\documentclass[a4paper,twocolumn,pre,showpacs]{revtex4}

\usepackage{amsmath,amssymb,amsfonts}
\usepackage{graphicx}
\usepackage{color}

\newcommand{\ham}{\mathcal{H}}
\newcommand{\saf}[1]{\textsf{#1}}
\newcommand{\nocc[2]}{n_{\saf{#1},{#2}}}

\begin{document}

\title{Hydration of an apolar solute in a two-dimensional waterlike
lattice fluid}

\author{C.~Buzano, E.~{De~Stefanis}, and M.~Pretti}

\affiliation{Istituto Nazionale per la Fisica della Materia (INFM)
and Dipartimento di Fisica, \\ Politecnico di Torino, Corso Duca
degli Abruzzi 24, I-10129 Torino, Italy}

\date{\today}

\begin{abstract}
In a previous work, we investigated a two-dimensional
lattice-fluid model, displaying some waterlike thermodynamic
anomalies. The model, defined on a triangular lattice, is now
extended to aqueous solutions with apolar species. Water molecules
are of the ``Mercedes Benz'' type, i.e., they possess a $D_3$
(equilateral triangle) symmetry, with three equivalent bonding
arms. Bond formation depends both on orientation and local
density. The insertion of inert molecules displays typical
signatures of hydrophobic hydration: large positive transfer free
energy, large negative transfer entropy (at low temperature),
strong temperature dependence of the transfer enthalpy and
entropy, i.e., large (positive) transfer heat capacity. Model
properties are derived by a generalized first order approximation
on a triangle cluster.
\end{abstract}


\maketitle

\section{Introduction}

The term hydrophobicity~\cite{DillScience1990} refers to peculiar
thermodynamic properties of the transfer process of an apolar
solute into water. In such a process, one generally observes large
positive transfer free energy, large negative transfer entropy (at
room temperature), and strong temperature dependence of the
transfer enthalpy and entropy, i.e, large (positive) transfer heat
capacity. Such anomalous behavior seems to play a central role for
relevant phenomena taking place in aqueous solutions, such as
folding of macromolecules and proteins, and formation of micelles
and membranes~\cite{Tanford1980}. For example, bio-polymers such
as proteins contain a significant fraction of apolar chemical
groups, and it is well established that the effective attraction
between apolar groups, due to hydrophobicity, gives an important
contribution both to the folding process and to stabilization of
the folded protein.

Despite several decades of research, the theory of the hydrophobic
effect is still incomplete. Different theoretical models have been
proposed to explain the anomalous behavior of water itself and
hydrophobic interactions, which have been recognized to be closely
related. A straightforward approach relies on the application of
molecular dynamics or Monte Carlo simulation methods to models
with more or less realistic three-dimensional water
geometry~\cite{StillingerRahman1974,Jorgensen1983,MahoneyJorgensen2000,Stanley2002,Paschek2004}.
This approach is powerful, but has some limitations. First of all,
large computational effort is needed, and properties involving
multiple derivatives of the free energy (such as transfer heat
capacity, whose behavior is a fingerprint of hydrophobicity)
cannot be determined easily. Moreover, the amount of different
interactions and geometric parameters included in the model makes
it difficult to find out relevant physical mechanisms which
determine observable properties.

A complementary approach involves investigation of simplified
models~\cite{SouthallDillHaymet2002,AshbaughTruskettDebenedetti2002,WidomBhimalapuramKoga2003},
with fewer parameters, geometric details, and degrees of freedom.
Such models should better allow to trace connections between
microscopic structure and observed properties, while the latter
can be usually analyzed in full detail, and in a large range of
thermodynamic conditions, with relatively small computational
effort. One of these attempts is based on the application of
scaled-particle theory~\cite{ReissFrischLebowitz1965} to
hydrophobic hydration~\cite{Lee1985,Lee1991}. These studies
suggest that the hydrophobic effect results mostly from the small
size of water molecules, and not from water structuring by the
solute. A recent and interesting descendant of scaled-particle
theory is the information theory approach by Pratt and
coworkers~\cite{Pratt2002,Garde1996}, based on previous knowledge
of water properties, such as the pair correlation function,
obtainable either by experiments or by simulations. The latter
approach suggests that water structuring induced by the solute,
though existing, may be scarcely relevant for a description of the
hydrophobic effect. The simplified molecular thermodynamic theory
of Ref.~\onlinecite{AshbaughTruskettDebenedetti2002} is
essentially in agreement with this conclusion. On the contrary,
different theories stress that the large positive heat capacity
variation, observed upon insertion of apolar solutes into water,
can only arise from a cooperative phenomenon, that is from induced
ordering of water molecules, so that a theory of the hydrophobic
effect should be based on a description of this phenomenon. It can
be observed that, at room temperature, hydration of apolar solutes
is energetically favorable but sufficiently unfavorable
entropically, with a resulting increased free energy. A simplified
way to reproduce this effect is for instance the one-dimensional
lattice model by Kolomeisky and Widom~\cite{KolomeiskyWidom1999},
extended also to two and three dimensions~\cite{BarkemaWidom2000}.
In that case, entropy penalty is achieved by lowering the large
number of possible orientations of a water molecule, if it has to
accommodate a neighboring hydrophobic solute. Another possibility
is to give water molecules a geometric structure, as in the
two-dimensional Mercedes Benz (MB) model, first introduced by
Ben-Naim in 1971~\cite{BenNaim1971}. In the latter model, water
molecules possess three equivalent bonding arms arranged as in the
Mercedes logo, and a geometric constraint (arm alignment) is
required for bond formation. An MB model has been investigated
quite recently by Dill and coworkers by means of several different
methods: Constant pressure Monte Carlo
simulations~\cite{SilversteinHaymetDill1998,SouthallDill2000},
entropy expansion~\cite{SilversteinDillHaymet2001}, and integral
equation
theory~\cite{UrbicVlachyKalyuzhnyiSouthallDill2000,UrbicVlachyKalyuzhnyiSouthallDill2002,UrbicVlachyKalyuzhnyiDill2003}.
A coherent picture of the hydrophobic effect phenomenology has
been worked out, allowing to obtain a microscopic view of several
anomalous properties of water both as a pure substance and as a
solvent. The latter studies follow the previously mentioned idea
of simplified models, that nevertheless are based on well defined
microscopic interactions, that is, on an energy function, without
previous knowledge of water properties. One important reason to do
so is the need of modelling water in a computationally convenient
way, in order to investigations on complex systems such as
biomolecules, for which water plays a key role.

According to the same idea, we have recently
investigated~\cite{BuzanoDestefanisPelizzolaPretti2004} a
lattice-fluid model of MB type on a (two-dimensional) triangular
lattice, with the aim of reproducing qualitatively the
thermodynamic anomalies of pure water. Of course, the lattice
allows important simplifications, so that a sufficiently accurate
modelling of water on a lattice may be quite an interesting issue.
Our model hamiltonian includes Van der Waals interaction and
hydrogen bonding, when two nearest neighbor MB water molecules
point an arm to each other. Bonds can be weakened by the presence
of a third competing molecule, close to the formed bond, to mimic
the fact that, if water molecules are too close to one another,
hydrogen bonds may be perturbed or broken. As far as bonding
properties are concerned, the model is equivalent to an early
model proposed by Bell and Lavis~\cite{BellLavisII1970}, but the
weakening criterion is different, that is, equivalent to the one
employed in quite a recent investigation by Patrykiejew and
coworkers~\cite{PatrykiejewPizioSokolowski1999,BruscoliniPelizzolaCasetti2002}.
Nevertheless, in the latter analysis the possibility of nonbonding
orientations for water molecules is neglected, and resemblance
with real water behavior is poor. In
Ref.~\cite{BuzanoDestefanisPelizzolaPretti2004} we observed that,
introducing such additional orientations, which account for
directionality of hydrogen bonds, the lattice model describes
several anomalous properties of pure water in a qualitatively
correct way. Here we extend the model to the case of an aqueous
solution, working out solvation thermodynamics for an inert
(apolar) solute. The analysis is also extended to transfer
properties of water in its own pure liquid. Our purpose is to
verify whether this simple model, which nevertheless accounts for
the competition between Van der Waals interactions and hydrogen
bonding in almost the same way as the off-lattice MB model, may be
able to reproduce at least the main features of hydrophobicity. In
particular, we also investigate how the model describes the
solvation process at a microscopic level, by comparing the average
number of hydrogen bonds formed by water molecules (hydrogen bond
coordination) in the bulk or in the hydration shell. We shall
carry out the analysis by means of a generalized first-order
approximation on a triangle cluster, which has been verified to be
quite accurate for the pure water
model~\cite{BuzanoDestefanisPelizzolaPretti2004}.

The paper is organized as follows. In Sec.~II we define the model
and recall the first-order approximation. In Sec~III we introduce
the thermodynamic quantities used to characterize the solvation
process, and compute them in the framework of the model. In
particular, we analyze transfer quantities for an inert molecule,
comparing them to the case in which hydrogen bonding interaction
is ``turned off''. Model predictions about solvation of water in
its own pure liquid are also reported. In Sec.~IV we investigate
hydrogen bond coordination, drawing a comparison with the results
of the off-lattice MB model. Sec.~V contains some concluding
remarks.

\section{The model and the first order approximation}

The model is defined on a two dimensional triangular lattice. A
lattice site can be empty or occupied by molecules of two
different chemical species, water (\saf{w}) or solute (\saf{s}). A
water molecule has three equivalent bonding arms, separated by
$2\pi/3$ angles, whereas a solute molecule is assumed to have no
internal structure. Two nearest-neighbor (NN) molecules of species
$\saf{x}$, $\saf{y}$ (with $\saf{x},\:\saf{y}\: =\:
\saf{w},\:\saf{s}$) interact with an attractive energy
$-\epsilon_{\saf{xy}}<0$, representing ordinary Van der Waals
forces. Moreover, if two arms of two NN water molecules point to
each other, an energy term $-\eta<0$ is added, to mimic the
formation of a hydrogen (H) bond. Due to the lattice structure, a
water molecule can form $3$ bonds at most, and has only 2 bonding
orientations, when its arms are aligned with the lattice. We also
assume that $w$ nonbonding configurations exist, where $w$ is a
model parameter, related to the H~bond breaking entropy. The
H~bond is weakened by an energy term $c_\saf{x}\eta/2$ ($c_\saf{x}
\in [0,1]$) when a third molecule of $\saf{x}$ species is on a
site near a formed bond. In the triangular lattice there are $2$
such weakening sites per bond, so that, when both are occupied by
$\saf{x}$~molecules, the H~bond contributes an energy $-(1 -
c_\saf{x})\eta$. The resulting water-solute interaction has two
components: The nonorientational Van der Waals term
$-\epsilon_{\saf{ws}}$ and the weakening term $c_\saf{s} \eta/2$.
The latter, which is an effective 3~body interaction, can be
viewed as a perturbation effect of the solute on a H~bond between
two water molecules.

Let us write the model hamiltonian. In order to introduce the
first order approximation, it is convenient to express it as a sum
over elementary triangles:
\begin{equation}
  \ham =
  \frac{1}{2} \sum_{\langle r,r',r'' \rangle}
  \ham_{i_r {i_r}_{'} {i_r}_{''}}
  ,
  \label{eq:ham}
\end{equation}
where $\ham_{ijk}$ is a contribution which will be referred to as
triangle hamiltonian, and $i_r,{i_r}_{'},{i_r}_{''}$ label site
configurations for the 3 vertices $r,r',r''$, respectively.
Possible site configurations are (see Tab.~\ref{tab-site-conf}):
``empty'' ($i=0$), ``bonding water'' (site occupied by a water
molecule in one of the $2$ orientations which can form bonds:
$i=1,2$), ``nonbonding water'' (site occupied by a water molecule
in one of the $w$ orientations which cannot form bonds: $i=3$),
``solute'' (site occupied by a solute molecule: $i=4$).
\begin{table}
  \caption{
    Possible site configurations,
    with corresponding labels~($i$), multiplicities~($w_i$), and
    occupation numbers for water~($\nocc[w]{i}$) and solute~($\nocc[s]{i}$).
  }
  \begin{ruledtabular}
  \begin{tabular}{l|ccccc}
    config. & empty & \includegraphics{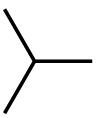} & \includegraphics{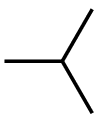} & \includegraphics{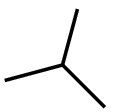} & \includegraphics{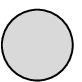}
    \cr
    \hline
    $i$ & 0 & 1 & 2 & 3 & 4\cr
    \hline
    $w_i$ & 1 & 1 & 1 & $w$ & 1 \cr
    \hline
    $\nocc[w]{i}$ & 0 & 1 & 1 & 1 & 0 \cr
    $\nocc[s]{i}$ & 0 & 0 & 0 & 0 & 1 \cr
  \end{tabular}
  \end{ruledtabular}
  \label{tab-site-conf}
\end{table}
Even if all configurations have unit multiplicity, except $i=3$,
it is convenient to introduce a generic multiplicity parameter
$w_i$, defined as in Tab.~\ref{tab-site-conf}. The triangle
hamiltonian can be written as
\begin{equation}
  \ham_{ijk} = H_{ijk} + H_{jki} + H_{kij}
  ,
  \label{eq:triham}
\end{equation}
where
\begin{equation}
  H_{ijk} = -
  \epsilon_\saf{xy} \nocc[x]{i} \nocc[y]{j}
  - \eta h_{ij} \left( 1 -
  c_\saf{x} \nocc[x]{k} \right)
  ,
  \label{eq:triham2}
\end{equation}
$\nocc[x]{i}$ is an occupation variable for the $\saf{x}$ species,
defined as in Tab.~\ref{tab-site-conf}, while $h_{ij}=1$ if the
pair configuration $(i,j)$ forms a H~bond, and $h_{ij}=0$
otherwise. Conventionally, repeated $\saf{x}$ and $\saf{y}$
indices are summed over their possible values $\saf{w},\saf{s}$.
Let us notice that triangle vertices are set on three triangular
sublattices, say $A,B,C$, and $i,j,k$ are assumed to denote
configurations of sites placed on $A,B,C$ sublattices
respectively. Assuming also that $A,B,C$ are ordered
counterclockwise on up-pointing triangles (whence clockwise on
down-pointing triangles), we can define $h_{ij}=1$ if $i=1$ and
$j=2$ and $h_{ij}=0$ otherwise. Let us also notice that both Van
der Waals ($-\epsilon_\saf{xy} \nocc[x]{i} \nocc[y]{j}$) and
H~bond energies ($-\eta h_{ij}$), that are 2-body terms, are split
between two triangles, whence the $1/2$~prefactor in
Eq.~\eqref{eq:ham}. On the contrary the 3-body weakening terms
($\eta h_{ij} c_\saf{x} \nocc[x]{k}/2$) are associated each one to
a given triangle, and the $1/2$ factor is absorbed in the
prefactor. Let us denote the triangle configuration probability by
$p_{ijk}$, and assume that the probability distribution is equal
for every triangle (no distinction between up- or down-pointing
triangles). Taking into account that there are 2 triangles per
site, we can write the following expression for the internal
energy per site of an infinite lattice
\begin{equation}
  u = \sum_{i=0}^4 \sum_{j=0}^4 \sum_{k=0}^4 w_i w_j w_k p_{ijk} \ham_{ijk}
  .
  \label{eq:intenergy}
\end{equation}
The multiplicity for the triangle configuration $(i,j,k)$ is given
by $w_i w_j w_k$, where $w_i$ is the previously mentioned
multiplicity parameter.

The details of the finite temperature analysis of the model, by a
generalized first order approximation on a triangle cluster,
follow the previous
paper~\cite{BuzanoDestefanisPelizzolaPretti2004}. Not being
interested in symmetry broken phases (ice), we introduce a
homogeneity condition since the beginning, after which
generalization is trivial. The grand-canonical free energy per
site $\omega = u - \mu_\saf{x} \rho_\saf{x} - Ts$ ($\mu_\saf{x}$
and $\rho_\saf{x}$ being respectively the chemical potential and
the density, or site occupation probability, for the $\saf{x}$
species, $T$ and $s$ being respectively the temperature and the
entropy per site), can be written as a functional in the triangle
probability distribution as
\begin{equation}
  \begin{split}
  \omega = &
  \sum_{i=0}^4 \sum_{j=0}^4 \sum_{k=0}^4 w_i w_j w_k p_{ijk}
  \left( \tilde{\ham}_{ijk} + T \ln p_{ijk} \right) \\
  & -2T \sum_{i=0}^4 w_i p_{i} \ln p_{i}
  ,
  \end{split}
  \label{eq:func}
\end{equation}
where $T$ is expressed in energy units (whence entropy in natural
units). It is noteworthy that the only variational parameter
in~$\omega$ is the triangle probability distribution, that is the
125 variables $\{p_{ijk}\}$, because $p_i$ (the site probability)
can be expressed as a marginal. According to the homogeneity
hypothesis, we can use the symmetrized expression
\begin{equation}
  p_i = \sum_{j=0}^4 \sum_{k=0}^4 w_j w_k
  \frac{p_{ijk}+p_{jki}+p_{kij}}{3}
  .
\end{equation}
The latter turns out to be convenient, in that it gives rise to
iterative (fixed point) equations which preserve homogeneity. The
energy part in Eq.~\eqref{eq:func} is given by the modified
hamiltonian $\tilde{\ham}_{ijk}$, defined as
\begin{equation}
  \tilde{\ham}_{ijk} = \ham_{ijk}
  - \mu_\saf{x} \frac{\nocc[x]{i} + \nocc[x]{j} + \nocc[x]{k}}{3}
  \label{eq:trihamtilde}
  .
\end{equation}
The minimization of~$\omega$ with respect to $\{p_{ijk}\}$, with
the normalization constraint, yields the equations
\begin{equation}
  p_{ijk} = \xi^{-1} e^{-\tilde{\ham}_{ijk}/T}
  {\left( p_i p_j p_k \right)}^{2/3}
  ,
  \label{eq:cvmeq}
\end{equation}
where $\xi$ is a normalization constant. Such equations can be
solved numerically by simple iteration (natural iteration
method~\cite{Kikuchi1974}). As a result, we obtain the triangle
probability values at equilibrium, from which one can compute the
thermal average of every observable. The substitution into
Eqs.~\eqref{eq:intenergy} and \eqref{eq:func} gives respectively
the equilibrium internal energy and free energy, and, by the way,
it is possible to show that
\begin{equation}
  \omega = -T \ln \xi
  .
\end{equation}

\section{Solvation thermodynamics}

Let us first introduce the thermodynamic quantities, generally
used to describe solvation from a macroscopic point of view, which
we shall evaluate for our model in the following. The physical
process we are interested in is the transfer of a component
($\saf{x} =\: \saf{w},\:\saf{s}$) into a water-solute mixture,
with solute density tending to zero. According to the Ben-Naim
standard~\cite{BenNaim1987}, this process is well characterized by
variation of the pseudo-chemical potential $\mu^{*}_\saf{x}$,
related to the ordinary chemical potential $\mu_\saf{x}$ of the
given component by
\begin{equation}
  \mu_\saf{x} = \mu^{*}_\saf{x} + T\log\rho_\saf{x}
  .
\end{equation}
Let us notice that, in the latter term on the right hand side, the
momentum partition function is missing, due to the fact that we
are dealing with a lattice model~\cite{BenNaim1987}. The solvation
free energy per molecule $\Delta g^*_\saf{x}$ is defined as the
difference between pseudo-chemical potential values of a molecule
$\saf{x}$ in the ideal gas phase ($g$) and in the liquid phase
($l$). For practical purposes, the differences between ideal and
real gas can be generally neglected~\cite{BenNaim1987}. For a
molecule of species $\saf{x}$ we then have
\begin{equation}
  \Delta g^*_\saf{x} = \mu^{*(l)}_\saf{x} - \mu^{*(g)}_\saf{x}
  \label{eq:mudg}
\end{equation}
where $\mu^{*(l)}_\saf{x}$ and $\mu^{*(g)}_\saf{x}$ are
pseudo-chemical potentials of $\saf{x}$ in the liquid and gas
phases, respectively. Now, if liquid and gas phase coexist in
equilibrium, as usual in experiments, the ordinary chemical
potentials of $\saf{x}$ in both phases must be equal, and by
simple algebra we obtain
\begin{equation}
  \Delta g^*_\saf{x} = -T \ln
  \frac{\rho_\saf{x}^{(l)}}{\rho_\saf{x}^{(g)}}
  ,
  \label{eq:realdg}
\end{equation}
where $\rho_\saf{x}^{(l)}$ and $\rho_\saf{x}^{(g)}$ are
respectively the $\saf{x}$ species densities in the liquid and in
the gas. Derived quantities, of interest in experiments, are the
solvation entropy
\begin{equation}
  \Delta s^*_\saf{x} = - \frac{\partial \Delta g^*_\saf{x}}
  {\partial T}\biggl\lvert_P
  ,
  \label{eq:constpentropy}
\end{equation}
the solvation enthalpy
\begin{equation}
  \Delta h^*_\saf{x} = \Delta g^*_\saf{x} + T \Delta s^*_\saf{x}
  ,
\end{equation}
and the solvation heat-capacity
\begin{equation}
  \Delta {c_P}^*_\saf{x} = \frac{\partial\Delta h^*_\saf{x}}
  {\partial T}\biggl\lvert_P
  .
\end{equation}
Let us notice that, in principle, we should distinguish between a
constant pressure derivative (as stated by definition) and a
temperature derivative taken along the liquid-vapor equilibrium
curve. In particular, we could not even use Eq.~\eqref{eq:realdg},
because we would move out of the equilibrium curve, at which the
ordinary chemical potentials are equal. Nevertheless, numerical
results~\cite{BenNaim1987} show that the difference is negligible,
definitely from a qualitative point of view. Therefore, we shall
always use the latter temperature derivative definition in our
calculations.

\begin{figure*}
  \resizebox{160mm}{!}{\includegraphics*[18mm,16mm][145mm,105mm]{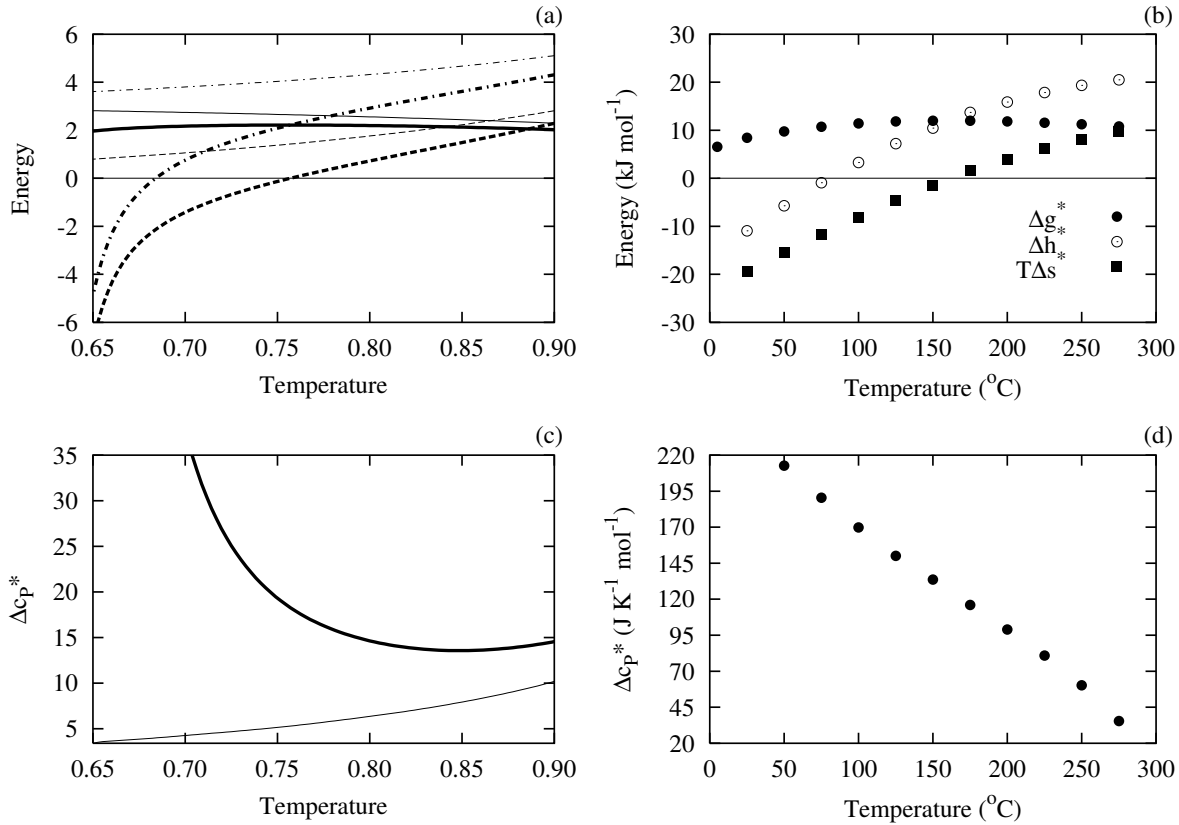}}
  \caption{
    (a) Solvation energies ($E/\epsilon_\saf{ww}$)
    vs temperature ($T/\epsilon_\saf{ww}$)
    for an ideal inert molecule in water
    at liquid-vapor coexistence:
    $E = \Delta g^*_\saf{s}$ (solid line),
    $E = T\Delta s^*_\saf{s}$ (dashed line),
    and $E = \Delta h^*_\saf{s}$ (dashed-dotted line).
    (b) Corresponding experimental data
    for transfer of argon into water~\cite{BenNaim1987}.
    (c) Solvation heat capacity ($\Delta {c_P}^*$)
    vs temperature ($T/\epsilon_\saf{ww}$)
    for an ideal inert molecule in water.
    (d) Corresponding experimental data
    for transfer of argon into water~\cite{BenNaim1987}.
    Thin lines in (a) and (c) denote solvation energies
    in nonbonding water ($\eta = 0$).
  }
  \label{fig:solute-water}
\end{figure*}

Let us start studying solvation of an inert molecule in water. We
set water parameters as follows: $\eta/\epsilon_{\saf{ww}} = 3$,
$c_{\saf{w}} = 0.8$, and $w = 20$. As shown in our previous
analysis~\cite{BuzanoDestefanisPelizzolaPretti2004}, this choice
corresponds to a waterlike behavior, with a liquid-vapor critical
point at $T/\epsilon_{\saf{ww}} \approx 1.18$, and with a
temperature of maximum density for the liquid around
$T/\epsilon_{\saf{ww}} \approx 0.67$ at low pressure. Solvation
thermodynamics concepts introduced above are independent of
density of components in the mixture. We choose to let solute
density assume very low values with respect to water density
(dilute solution limit), in order to compare with experiments. The
``inert'' nature of the solute is described, in the model
framework, by setting to zero the solute-solute
($\epsilon_{\saf{ss}}$) and water-solute ($\epsilon_{\saf{ws}}$)
Van der Waals interaction energies. At the moment, we also set the
weakening parameter $c_\saf{s}$ to zero, assuming that the ideally
inert solute does not weaken H~bonds in its neighborhood. The
effect of nonzero values for this parameter, which may describe
for instance a volume interaction, resulting in a perturbation of
the geometry (and hence of the energy) of H~bonds, will be taken
into account later. The temperature trends of the free energy,
entropy, and enthalpy of transfer are given in
Fig.~\ref{fig:solute-water}(a); the transfer heat capacity in
Fig.~\ref{fig:solute-water}(c). In order to compare with
experimental data~\cite{BenNaim1987}, all quantities are evaluated
at liquid-vapor coexistence. For the perfectly inert solute, we
have verified that concentration does not affect the results at
all. To represent roughly the experimental temperature range
(between $0^\circ\,\mathrm{C}$ and $300^\circ\,\mathrm{C}$) we
report model results between $T/\epsilon_{\saf{ww}} = 0.65$ (just
below the temperature of maximum density for pure liquid water)
and $T/\epsilon_{\saf{ww}} = 0.90$ (about half way between the
previous temperature and the critical temperature). Remarkably, it
turns out that the model, despite its simplicity, displays the
defining feature of the hydrophobic effect: the solvation free
energy is positive and large, while the solvation entropy is
negative at low temperatures and becomes positive upon increasing
temperature; the heat capacity is positive and large, and also the
decreasing trend with temperature is essentially reproduced. The
increasing trend at high temperature is related to the the fact
that we are approaching the liquid-vapor critical point. Negative
solvation entropy at low (room) temperature is a clear indication
that solute insertion into the mixture orders the system. The
corresponding positive (unfavorable) contribution to free energy
compensates a negative (favorable) enthalpic contribution, giving
rise to a positive solvation free energy. At higher temperature,
enthalpic and entropic contributions change sign, but they still
have the same trend of compensating each other. As observed in
experiments, the model predicts two different temperatures $T_H$
and $T_S$ at which the transfer enthalpy and entropy vanish,
respectively [see Fig.~\ref{fig:solute-water}(b)]. This behavior
is to be ascribed to the thermodynamics of H~bonding and, in order
to rationalize this fact in the model framework, let us also
analyze transfer quantities for the case $\eta=0$, i.e., turning
off H~bond interactions [see Fig.~\ref{fig:solute-water}(a),(c)].
As expected, the results are similar in the high temperature
regime, where there is a high probability that hydrogen bonds are
broken by thermal fluctuations, whereas they change more and more
dramatically upon decreasing temperature, and in particular the
regions of negative transfer entropy and enthalpy completely
disappear. This facts confirm that H~bonding is the key element
for system ordering, upon insertion of an inert molecule.
Accordingly, also the divergentlike trend of the heat capacity
upon decreasing temperature (related to the fact that the liquid
phase is approaching a stability
limit~\cite{BuzanoDestefanisPelizzolaPretti2004}) is suppressed.
The process is now completely dominated by the transfer enthalpy,
with a large and positive transfer free energy, and a positive
transfer entropy. The transfer quantities behave qualitatively as
those observed in solvation experiments of noble gas molecules in
ordinary liquids~\cite{DaviesDuncan1967,BenNaim1987}, and are
relatively independent of temperature. In fact, with $\eta=0$, a
water molecule can be viewed as a nonpolar molecule with Van der
Waals interaction energy~$\epsilon_{\saf{ww}}$.

\begin{figure*}
  \resizebox{160mm}{!}{\includegraphics*[18mm,16mm][145mm,62mm]{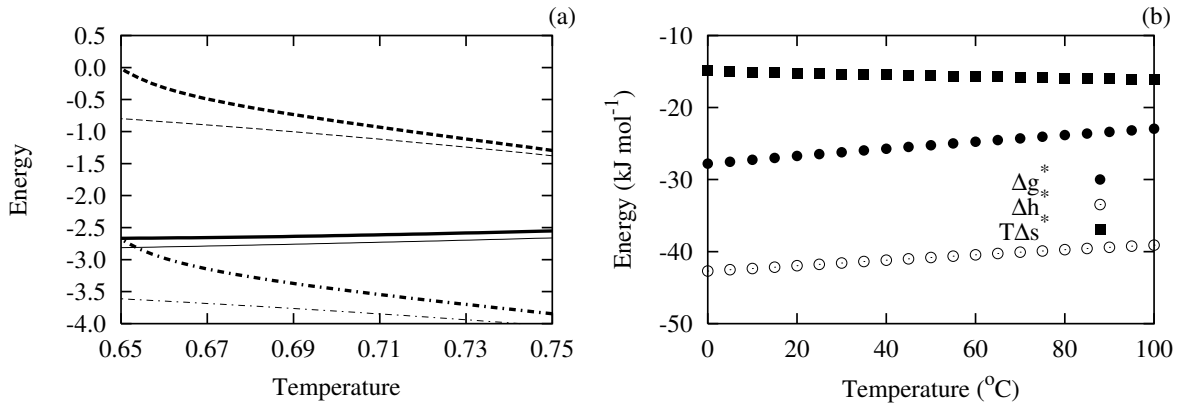}}
  \caption{
    (a) Solvation energies ($E/\epsilon_\saf{ww}$)
    vs temperature ($T/\epsilon_\saf{ww}$)
    for a water molecule into pure liquid water
    at liquid-vapor coexistence:
    $E = \Delta g^*_\saf{s}$ (solid line),
    $E = T\Delta s^*_\saf{s}$ (dashed line),
    and $E = \Delta h^*_\saf{s}$ (dashed-dotted line).
    (b) Corresponding experimental data~\cite{BenNaim1987}.
    Thin lines in (a) denote solvation energies
    for nonbonding water ($\eta = 0$).
  }
  \label{fig:water-water}
\end{figure*}
Let us now consider also the solvation of water in its own pure
liquid. The corresponding transfer energies obtained by the model
are displayed in Fig.~\ref{fig:water-water}(a), where we have
reduced the temperature interval, in order to compare with
available experimental results~\cite{BenNaim1987}, reported in
Fig.~\ref{fig:water-water}(b). In contrast to the inert molecule
case, here the absolute values of solvation free energy and
entropy are considerably lower and the enthalpy, rather than the
entropy, dominates the solvation process, and all quantities are
relatively independent of temperature. These features characterize
a regular transfer process, like the solvation of an ordinary
fluid molecule from a gas phase into its pure liquid phase. In
this case, upon removing H bond interactions [thin lines in
Fig.~\ref{fig:water-water}(a)], very little changes are observed
in the solvation energies, except at very low temperature, where
we are approaching the stability limit for liquid
water~\cite{BuzanoDestefanisPelizzolaPretti2004}. Let us discuss
two issues about these results. First, the fact that so little
changes are caused by turning on or off H~bonds can be
rationalized on the basis of the microscopic model interactions.
The insertion of a water molecule into pure liquid water should
imply in principle the formation of new H bonds, but the model is
such that insertion of a new water molecule also weakens other H
bonds in its neighborhood, and the two effects nearly compensate
each other. Second, let us notice that solvation enthalpy
decreases upon increasing temperature, that is, the solvation heat
capacity is negative, in contrast with experiments. We do not have
an explanation for this fact, but we can observe that an analogous
effect is observed when the model is reduced to describe a regular
solvation process, that is when H bonds are turned off. This
suggest that there is probably a limitation of the lattice
description, that anyway has nothing to do with the peculiarities
of water. The effect is quantitatively small, so that it is hidden
by other large enthalpic and entropic effects observed in the case
of hydrophobic solvation.

\section{Hydrogen bond coordination}

\begin{figure}
  \begin{picture}(0,0)%
  \includegraphics{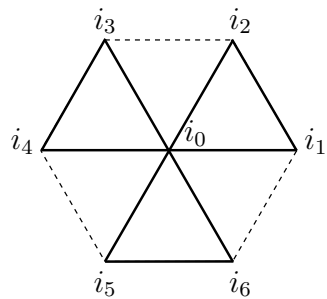}%
  \end{picture}%
  \setlength{\unitlength}{4144sp}%
  \begingroup\makeatletter\ifx\SetFigFont\undefined%
  \gdef\SetFigFont#1#2#3#4#5{%
    \reset@font\fontsize{#1}{#2pt}%
    \fontfamily{#3}\fontseries{#4}\fontshape{#5}%
    \selectfont}%
  \fi\endgroup%
  \begin{picture}(2170,1789)(-179,-839)
  \put(316,794){\makebox(0,0)[lb]{\smash{{\SetFigFont{12}{14.4}{\familydefault}{\mddefault}{\updefault}{\color[rgb]{0,0,0}$i_3$}%
  }}}}
  \put(316,-781){\makebox(0,0)[lb]{\smash{{\SetFigFont{12}{14.4}{\familydefault}{\mddefault}{\updefault}{\color[rgb]{0,0,0}$i_5$}%
  }}}}
  \put(1576, 74){\makebox(0,0)[lb]{\smash{{\SetFigFont{12}{14.4}{\familydefault}{\mddefault}{\updefault}{\color[rgb]{0,0,0}$i_1$}%
  }}}}
  \put(856,119){\makebox(0,0)[lb]{\smash{{\SetFigFont{12}{14.4}{\familydefault}{\mddefault}{\updefault}{\color[rgb]{0,0,0}$i_0$}%
  }}}}
  \put(-179, 74){\makebox(0,0)[lb]{\smash{{\SetFigFont{12}{14.4}{\familydefault}{\mddefault}{\updefault}{\color[rgb]{0,0,0}  $i_4$}%
  }}}}
  \put(1126,-781){\makebox(0,0)[lb]{\smash{{\SetFigFont{12}{14.4}{\familydefault}{\mddefault}{\updefault}{\color[rgb]{0,0,0}$i_6$}%
  }}}}
  \put(1126,794){\makebox(0,0)[lb]{\smash{{\SetFigFont{12}{14.4}{\familydefault}{\mddefault}{\updefault}{\color[rgb]{0,0,0}$i_2$}%
  }}}}
  \end{picture}%
  \caption{
    Site labels for the hexagon probability distribution.
    Thick lines denote triangles for which
    correlations are taken into account.
  }
  \label{fig:hexagon}
\end{figure}

So far, we have always considered an ideal, perfectly inert
solute, setting to zero all interactions with water
($\epsilon_\saf{ws}$, $c_\saf{s}$) and with itself
($\epsilon_\saf{ss}$). Now we investigate the role of the
$c_\saf{s}$ parameter, which represents a weakening of H bonds,
induced by the presence of a solute molecule. Let us recall that,
in our model, the presence of too many water molecules close to
one another weakens the H bond strength, through the $c_\saf{w}$
parameter, to mimic the fact that too low average distance is
unfavorable for H bonds. On the contrary, a different (lower)
weakening parameter $c_\saf{s}$ for a solute molecule, can
represent a perturbation of H~bonds, due to an excluded volume
effect. Anyway, the solute weakening parameter $c_\saf{s}$ is a
way of tuning the degree of water ordering induced by the solute.
In order to characterize this effect, together with the role of
the $c_\saf{s}$ parameter, let us investigate the H bond
coordination, that is, the average number of hydrogen bonds per
molecule. In quite recent papers, Dill and
coworkers~\cite{SilversteinHaymetDill1998} suggested, on the basis
of their off-lattice MB model, that this parameter is the
appropriate one to rationalize the temperature~$T_S$ at which the
transfer entropy vanishes. In particular they distinguished
between H bond coordination for molecules in bulk water and in a
hydration shell. We can evaluate analogous parameters also for our
model, in the framework of the first order approximation. Each
water molecule can form bonds with NN molecules only, therefore it
is necessary to compute the joint probability distributions of a
given site with its 6 NNs (hexagon probability). According to the
Husimi lattice formulation of the first order approximation, it is
easy to see that only certain triangle correlations are taken into
account, as shown schematically in Fig~\ref{fig:hexagon}.
Therefore, $i_0$~being the central site configurations and
$i_1\dots\,i_6$ the NNs configurations, the hexagon probability
distribution reads
\begin{equation}
  p_{i_0 i_1 \dots \, i_6} =
  \frac{p_{i_0 i_1 i_2} p_{i_0 i_3 i_4} p_{i_0 i_5 i_6}}
  {{p_{i_0}}^2}
  .
  \label{eq:hexprob}
\end{equation}
We shall evaluate bond coordinations according to this
factorization, which, let us notice, is perfectly consistent with
our approximation scheme~\cite{Pretti2003}. Let us consider a
water molecule in a bonding configuration, for example $i=1$. It
is not necessary to consider also the other ($i=2$) bonding
configuration, due to the fact that the liquid phase is isotropic,
whereas we do not take into account nonbonding configurations
($i=3$), because in that case the probability of forming a bond is
zero. The bulk and shell coordinations can be written respectively
as
\begin{eqnarray}
  && {\langle q \rangle}_B = \sum_{q=1}^3 q P_{q|B} \\
  && {\langle q \rangle}_S = \sum_{q=1}^3 q P_{q|S}
\end{eqnarray}
where $P_{q|B}$, or $P_{q|S}$ respectively, is the probability
that the given molecule forms $q$ bonds, given that its NN sites
host no solute molecules (bulk water) or at least one solute
molecule (hydration shell). Working in the infinite dilution
limit, the probability of configurations with more than one NN
solute is actually a small corrections over the probability of
having just one solute molecule. Making use of the Bayes theorem,
the bulk conditional probability can be rewritten as
\begin{equation}
  P_{q|B} = \frac{P_{q,B}}{P_{B}}
  ,
\end{equation}
where $P_{q,B}$ is the probability that the central molecule forms
$q$ bonds, {\em and} that its NN sites host no solute molecules,
while $P_{B}$ is simply the probability that NNs host no solute
molecules. Moreover, making use also of the total probability
theorem, the shell conditional probability can be rewritten as
\begin{equation}
  P_{q|S} = \frac{P_{q,S}}{P_{S}} = \frac{P_{q}-P_{q,B}}{1-P_{B}}
  ,
\end{equation}
where $P_{q}$ is the total probability that the central molecule
forms $q$ bonds.

We can now evaluate the required probabilities, making use of the
factorization~\eqref{eq:hexprob}. First of all, it is easy to see
that
\begin{equation}
  P_{q} = {3 \choose q}
  \frac{{p_{12}}^{q} (p_1-p_{12})^{3-q}}
  {{p_{1}}^{3}}
  ,
\end{equation}
where
\begin{eqnarray}
  && p_{1} = \sum_{j=0}^4 \sum_{k=0}^4 w_j w_k p_{1jk} \\
  && p_{12} = \sum_{k=0}^4 w_k p_{12k}
  .
\end{eqnarray}
In fact, the central molecule, in the given $i=1$ configuration,
can form bonds along its $3$ arms. Therefore, ${p_{12}}^q$ is the
probability that a bond is formed along $q$ given arms, while
$(p_1-p_{12})^{3-q}$ is the probability that bond is not formed
along the remaining $3-q$ arms. The probability factorizes,
because arms lie on different triangles, which are uncorrelated,
according to Eq.~\eqref{eq:hexprob}. The binomial coefficient
accounts for different choices of $q$ bonds along $3$ arms, while
the denominator is due to the fact that $P_q$, though not
specified by the notation, is a conditional probability, with
respect to the presence of a central water molecule in the $i=1$
configuration. The joint probability $P_{q,B}$ can be evaluated in
similar way
\begin{equation}
  P_{q,B} = {3 \choose q}
  \frac{ \left. \tilde{p}_{12} \right.^q (\tilde{p}_1-\tilde{p}_{12})^{3-q}}
  {{p_{1}}^{3}}
  ,
\end{equation}
where we have to assume that no solute molecule is present in the
neighborhood, that is
\begin{eqnarray}
  && \tilde{p}_{1} = \sum_{j=0}^3 \sum_{k=0}^3 w_j w_k p_{1jk} \\
  && \tilde{p}_{12} = \sum_{k=0}^3 w_k p_{12k}
  = p_{12} - p_{124}
  .
\end{eqnarray}
Finally, the probability that no solute molecule is in the
neighborhood (bulk condition) can be written as
\begin{equation}
  P_{B} =
  \frac{\left.\tilde{p}_{1}\right.^3}{\left.p_{1}\right.^{3}}
  .
\end{equation}
\begin{figure}
  \resizebox{90mm}{!}{\includegraphics*{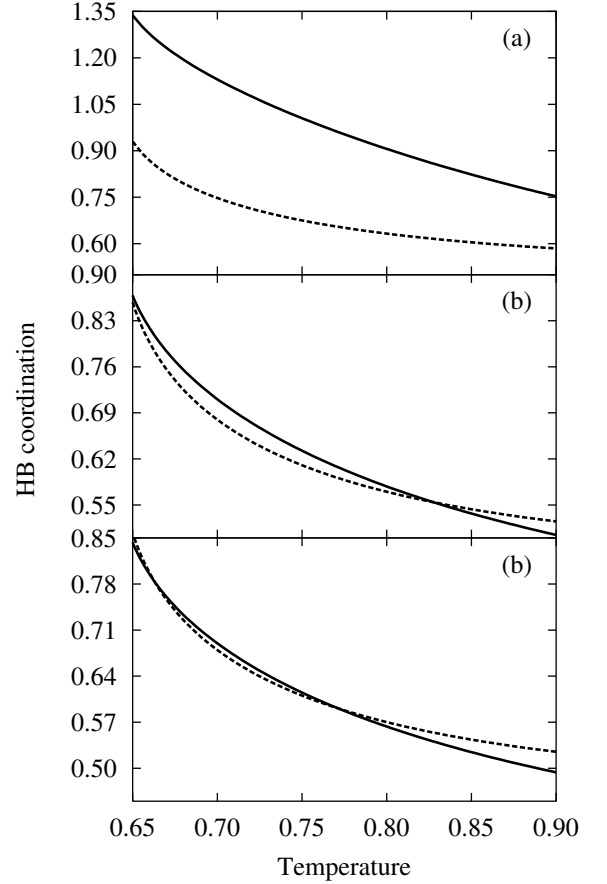}}
  \caption{H~bond coordination for a shell water molecule (solid line) and
    a bulk water molecule (dashed line) vs temperature.
    (a) $c_\saf{s} = 0$,
    (b) $c_\saf{s} = 0.37$, and
    (c) $c_\saf{s} = 0.405$.
  }
  \label{fig:hbcoord}
\end{figure}

\begin{figure}
  \resizebox{80mm}{!}{\includegraphics*{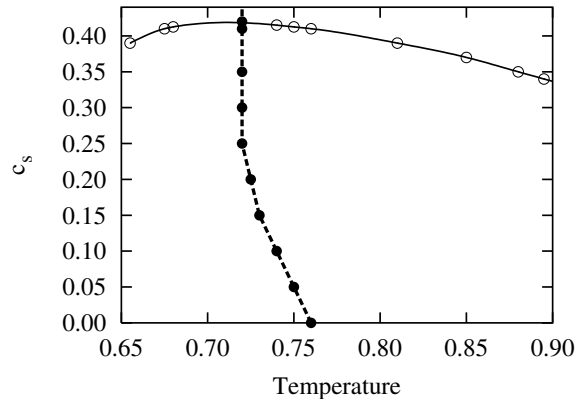}}
  \caption{Weakening parameter $c_\saf{s}$ vs crossing temperature
    $T_S'$, where ${\langle q \rangle}_S = {\langle q \rangle}_B$,
    (empty circles interpolated by a solid line)
    and vs $T_S$, where $\Delta s^*_\saf{s} = 0$,
    (filled circles interpolated by a dashed line).}
  \label{fig:ts_cs}
\end{figure}

In the work of Dill and
coworkers~\cite{SilversteinHaymetDill1998}, at low temperatures
H~bond coordination for shell water ${\langle q \rangle}_S$ is
greater than H~bond coordination for bulk water ${\langle q
\rangle}_B$. This behavior is reversed in the high temperature
region, where H~bonds are preferably formed by bulk water. In this
model, the crossing temperature $T_S'$ (where ${\langle q
\rangle}_S = {\langle q \rangle}_B$) is approximately equal to
$T_S$ (where $\Delta s^*_\saf{s} = 0$). As a consequence, average
H bond coordination seems to be strictly related to the transfer
entropy. Let us analyze what happens in our model. First of all,
we observe that the range of values of H bond coordination is
generally much lower than the maximum value of 3 H bonds, which
can be formed by a single water molecule (see
Fig.~\ref{fig:hbcoord}). This fact may be related to a peculiarity
of the lattice model, in which H bonds can be formed just along
given directions. Moreover, as far as the temperature behavior is
concerned, upon varying the weakening parameter $c_\saf{s}$, we
can have in principle four different scenarios. As shown in
Fig.~\ref{fig:hbcoord}(a), at low values of $c_\saf{s}$ (low H
bond weakening near a solute molecule), shell coordination
${\langle q \rangle}_S$ is greater than bulk coordination
${\langle q \rangle}_B$ for all temperatures. Increasing the
parameter, it is possible to observe a scenario similar to the one
found by Dill and coworkers: a crossing temperature $T_S'$ between
a low temperature region dominated by ${\langle q \rangle}_S$ and
a high temperature one dominated by ${\langle q \rangle}_B$
[Fig~\ref{fig:hbcoord}(b)]. Increasing $c_\saf{s}$ further, a more
complicated scenario, involving two crossing temperatures, can be
observed. The region where ${\langle q \rangle}_S$ is greater than
${\langle q \rangle}_B$ is upper-bounded by $T_S'$, as in the
previous scenario, and lower-bounded by a different crossing
temperature [Fig.~\ref{fig:hbcoord}(c)]. This means that the high
temperature behavior is restored at low temperature, though in a
region very close to the stability limit for the liquid phase,
which would be very difficult to reach in a real system. Finally,
for very high values of $c_\saf{s}$ (high weakening of H~bonds
near a solute molecule) a fourth scenario is observed, reversing
the behavior of the first scenario, that is, ${\langle q
\rangle}_B > {\langle q \rangle}_S$ for all temperatures. All
possible behaviors are summarized in Fig.~\ref{fig:ts_cs}, where
the crossing temperatures $T_S$ and $T_S'$ are displayed as
functions of the weakening parameter $c_\saf{s}$.
As a comparison, in the same figure, we have reported the zero
entropy temperature $T_S$, as a function of the same weakening
parameter $c_\saf{s}$. It can be easily observed that, while
H~bond coordination is strongly dependent on the weakening
parameter, the transfer entropy turns out to be a much more
``robust'' effect, and the $T_S$ temperature has only a relatively
slight dependence on the weakening parameter.

\section{Conclusions}

In this paper we have considered a two-dimensional waterlike
lattice model, defined on the triangular lattice, which we had
previously shown to reproduce qualitatively typical thermodynamic
anomalies of pure water, and extended the model to the case of an
aqueous solution. Water molecules are of the Mercedes Benz type,
with three equivalent bonding arms, while solute molecules have no
internal degrees of freedom. We have performed our calculations by
means of a generalized first order approximation on a triangle
cluster, which requires quite a small computational effort, and
had been shown to behave quite accurately for the pure water
model. In particular, in this paper we have addressed the issue of
dilute solutions of inert (apolar) solutes, that is, the
hydrophobic effect, and we have investigated thermodynamic
equilibrium between liquid and vapor, working out solvation
quantities in this case. It turns out that the model qualitatively
reproduces the peculiar features that are usually believed to be
the fingerprints of hydrophobicity. The solvation free energy is
positive, meaning that solvation is unfavorable, while the entropy
and enthalpy display an increasing trend and cross zero at two
different temperatures. The solvation heat capacity is large and
decreases upon increasing temperature. The model results compare
qualitatively well with experimental results about solvation of
noble gases into water. We have investigated the effect of
H~bonding, by comparing the previously mentioned results with
those obtained by setting to zero the H~bond energy parameter. In
this case, we have obtained transfer quantities that approach the
ones computed {\em with} H~bonds at high temperatures, but that
largely deviates from them upon decreasing temperatures, in the
region were H~bonding begins to dominate. In particular, we have
observed that, while the disaffinity of the solute for the solvent
remains, and is signaled by the positive large value of the
solvation free energy, such a disaffinity is mainly of enthalpic
nature, and both the enthalpy and entropy of solvation remain
positive at all observed temperatures, so that the strong
temperature dependence, which is typical of the hydrophobic
effect, disappears.

In order to a check of the model, we have also investigated
solvation of water into water at liquid-vapor equilibrium, for
which experimental data are available. We have found qualitative
agreement in the values of solvation free energy, entropy and
enthalpy, though there is some discrepancy in the temperature
dependence of enthalphy, which indicates a negative solvation heat
capacity, in contrast with experiments. Though not reporting
details in the paper, we have verified that this fact is neither
to be related to the approximation of the ideal/real gas phase,
nor with the temperature derivative approximation, mentioned in
Sec.~III. On the contrary, we have observed that the same kind of
discrepancy may be observed in the case of zero H~bond energy,
that is, for a regular solution. Therefore, we suspect that the
discrepancy is to be related to an intrinsic limitation of lattice
modelling. The effect is relatively small, so that it is
completely invisible, when the dominant effect of H~bonding is
turned on.

Finally, we have performed a calculation of the average number of
H~bonds formed by a single water molecule (H~bond coordination, in
short), in two different cases, namely when the molecule is placed
in the first hydration shell of a solute molecule (shell
coordination), and when it is not (bulk coordination). According
to Dill's Mercedes Benz model, these two parameters are clearly
related to the solvation entropy behavior, namely, negative
solvation entropies (low temperature) correspond to higher shell
coordination, while positive solvation entropies correspond to
higher bulk coordination. Due to the fact that our purpose was to
obtain a kind of lattice version of the MB model, describing
essentially the same physics underlying the observed
phenomenology, we have tried to verify whether the same effect
could be observed in our model or not. The answer is essentially
no, but some interesting observations can be done. We have
performed the investigation, upon varying the solute weakening
parameter, which, in our model, is a way of tuning the degree of
water ordering induced by the solute. We have observed that such
parameter strongly affects the behavior of H~bond coordination,
and in particular the temperature at which shell and bulk
coordinations become equal. On the contrary the temperature of
zero entropy $T_S$, which is actually one of the striking features
of the hydrophobic effect, is quite ``robust'' and relatively
independent of variations of the weakening parameter. Two
questions arise from the discussed behavior. On the one hand, we
may suspect that the lattice model is definitely too simple to
capture the microscopic physics of the hydrophobic effect in a
correct way, or that the approximation level is insufficient. But
the robustness of the zero entropy temperature may also suggest
that the simple relationship between the balance of bulk and shell
H~bonds and the zero of transfer entropy, proposed by Dill and
coworkers, may be model-dependent, and ought to be verified more
carefully. We plan to investigate on such issues in future works.


\end{document}